\documentclass[12pt,preprint]{aastex}

\newcounter{address}
\newcommand{\latin}[1]{{#1}}
\newcommand{\unit}[1]{\mathrm{#1}}

\newcommand{\eg}{\latin{e.g.}}
\newcommand{\bandz}[1]{{}^{0.3}{#1}}
\newcommand{\mean}[1]{\left<{#1}\right>}
\newcommand{\di}{\mathrm{d}}
\newcommand{\ggz}{\bandz{g}}
\newcommand{\rrz}{\bandz{r}}
\newcommand{\iiz}{\bandz{i}}
\newcommand{\wwsi}{w_{si}}
\newcommand{\rrp}{r_\mathrm{p}}
\newcommand{\rrf}{r_\mathrm{close}}
\newcommand{\ttmerge}{t_{\mathrm{merge},i}}
\newcommand{\ttorbit}{t_\mathrm{orbit}}
\newcommand{\ttdyn}{t_{\mathrm{dyn},i}}
\newcommand{\nns}{n_s}
\newcommand{\nni}{n_i}
\newcommand{\NNi}{N_i}
\newcommand{\MMg}{{}[M_{\ggz}-5\,\log_{10}h]}
\newcommand{\MMi}{{}[M_{\iiz}-5\,\log_{10}h]}
\newcommand{\set}[1]{\mathbb{#1}}
\newcommand{\setDDs}{\set{D}_s}
\newcommand{\setDDi}{\set{D}_i}
\newcommand{\setRRs}{\set{R}_s}
\newcommand{\setRRi}{\set{R}_i}
\newcommand{\DDs}{D_s}
\newcommand{\DDi}{D_i}
\newcommand{\RRs}{R_s}
\newcommand{\RRi}{R_i}
\newcommand{\xisi}{\xi_{si}}
\newcommand{\Lstar}{L^{\ast}}
\newcommand{\Mpc}{\unit{Mpc}}
\newcommand{\kpc}{\unit{kpc}}
\newcommand{\Gyr}{\unit{Gyr}}
\newcommand{\km}{\unit{km}}
\newcommand{\s}{\unit{s}}
\renewcommand{\arcsec}{\unit{arcsec}}
\newcommand{\percent}{\unit{percent}}
\renewcommand{\mag}{\unit{mag}}
\newcommand{\hMpc}{h^{-1}\,\Mpc}
\newcommand{\hkpc}{h^{-1}\,\kpc}

\begin{document}

\title{The growth of luminous red galaxies by merging}
\author{
	Morad~Masjedi\altaffilmark{\ref{NYU}},
	David~W.~Hogg\altaffilmark{\ref{NYU},\ref{email}},
	Michael~R.~Blanton\altaffilmark{\ref{NYU}}
}

\setcounter{address}{1}
\altaffiltext{\theaddress}{\stepcounter{address}\label{NYU} Center for
Cosmology and Particle Physics, Department of Physics, New York
University, 4 Washington Pl, New York, NY 10003}
\altaffiltext{\theaddress}{\stepcounter{address}\label{email} To whom
correspondence should be addressed: \texttt{david.hogg@nyu.edu}}

\begin{abstract}
We study the role of major and minor mergers in the mass growth of
luminous red galaxies.  We present small-scale ($0.01<r<8\,\hMpc$)
projected cross-correlation functions of $23043$ luminous early-type
galaxies from the Sloan Digital Sky Survey (SDSS) Luminous Red Galaxy
(LRG) sample ($0.16<z<0.30$, $\MMi\approx -22.75\,\mag$) with all
their companions in the SDSS imaging sample, split into color and
luminosity subsamples with $\MMi<-18\,\mag$.  We de-project the
two-dimensional functions to obtain three-dimensional real-space
LRG--galaxy cross-correlation functions for each companion
subsample. We find that the cross-correlation functions are not purely
power-law and that there is a clear ``one-halo'' to ``two-halo''
transition near $1\,\hMpc$.  We convert these results into close pair
statistics and estimate the LRG accretion rate from each companion
galaxy subsample using timescales from dynamical friction arguments
for each subsample of the companions.  We find that the accretion onto
LRGs is dominated by dry mergers of galaxies more luminous than
$\Lstar$.  We integrate the luminosity accretion rate from mergers
over all companion galaxy subsamples and find that LRGs are growing by
$[1.7\pm 0.1]$ percent per $\Gyr$, on average, from merger activity at
redshift $z\sim 0.25$.  This rate is almost certainly an over-estimate
because we have assumed that all close pairs are merging as quickly as
dynamical friction allows; nonetheless it is on the low side of the
panoply of measurements in the literature, and lower than any rate
predicted from theory.
\end{abstract}


\section{Introduction}

In the current paradigm for galaxy formation, the massive dark-matter
halos in which galaxies reside have assembled throughout cosmic time
by accretion and merging of smaller parts.  There remains, however,
substantial uncertainty about the formation and evolution of the
galaxies in those halos, and the evolution of the stars and gas that
compose them.  A particularly important galaxy subpopulation in this
research context is the luminous end of the red sequence of galaxies
--- consisting of concentrated, smooth, and typically elliptical
galaxies that preferentially live in the densest regions of the
Universe \citep{sandage72a, schneider83a, roberts94a, postman95a,
blanton03d}. Such galaxies account for a large fraction of the total
stellar mass in the Universe \citep{hogg02a, rudnick06a, brown07a},
and thus understanding their formation is critical to understanding
galaxy formation in general. This red sequence is clearly separated
from the blue sequence of galaxies, which are typically lower in mass,
more star-forming, gas-rich, morphologically ``spiral,'' and
preferentially populate isolated regions \citep{strateva01a,
blanton03d, baldry04a, balogh04b}.

With uniform spectra, deep absorption lines, highly clustered
distribution (``bias'' around 2), and high luminosities (absolute
magnitudes around $-23\,\mag$), the Luminous Red Galaxies (LRGs) are
excellent tracers of the density field on large scales
\citep{eisenstein01a}. For this reason, surveys of LRGs were the first
to conclusively demonstrate the homogeneity of the Universe at low
redshifts \citep{hogg05a} and to detect the baryon acoustic
oscillation feature in the correlation function
\citep{eisenstein05a}. Their continuing importance to understanding
fundamental cosmology underlines the need to better understand their
nature.

Because luminous red galaxies are typically supported by velocity
dispersion and not orderly rotation, it has been hypothesized that
they form from mergers of two or more smaller galaxies
\citep{toomre77a}.  Such mergers have been shown in numerical
simulations to produce disordered and velocity dispersion supported
systems not unlike observed ellipticals \citep{negroponte83a,
barnes96a, naab03a, cox06b}. In addition to such ``major'' mergers,
LRGs could in principle grow over time from an accumulation of smaller
mergers. If at least one of the merging galaxies is gas-rich, it often
shows a large star-formation rate \citep{barton00a, lambas03a,
nikolic04a, smith07a}, and indeed a significant fraction of the total
star-formation in the Universe may occur in such events. However,
there is also a population of mergers of two red galaxies, in which no
star-formation occurs, an event usually referred to as a ``dry
merger'' \citep{bell06b, vandokkum05a}.

A growing consensus of groups studying the high redshift Universe find
that the luminous red galaxies appear to grow in stellar mass over
time \citep{bell04a, willmer05a, blanton06a, wake06a, brown07a,
faber07a}. This growth can occur in several ways.  First, in principle
they may have ongoing star-formation.  This is very unlikely, since
the red galaxies show little signs of such star-formation.  Second,
luminous blue galaxies may transform to red galaxies.  This is also
unlikely at the very luminous end, since the number density of
luminous blue galaxies is far lower than that of red galaxies, even at
high redshift.  Third, and most likely, the luminous red galaxies may
grow through either major or minor mergers.

If the change in the galaxy population over time is to be explained at
least partly by mergers, we must be able to find these mergers before
or as they occur in appropriate numbers.  ``Instantaneous'' studies of
the merger rate are complementary to, and must be consistent with, the
global studies in the change of the galaxy population. To evaluate
whether mergers can explain these changes, the work presented here is
designed to measure the accretion or merger rate of companion galaxies
into LRGs at redshift $z\sim 0.25$.

It is also the case that in the CDM paradigm for structure formation,
galaxies reside in mass concentrations that are built from merging and
accretion of smaller concentrations over cosmic time.  It is
unavoidable that this merging in the dark sector is associated, at
some level, with merging of observable galaxies (\eg,
\citealt{murali02a, maller06a, conroy07a}).

There are many galaxy--galaxy merger rate estimates in the literature,
which involve identifying a class of pre-merger close pairs
\citep{carlberg94a, patton97a, vandokkum99a, carlberg00a, patton00a,
lin04a, masjedi06a, bell06a, depropris07a}, a class of post-merger
galaxies based on star-formation indicators \citep{quintero04a}, or a
class of currently merging sources based on disturbed or merging
morphologies \citep{abraham96, conselice03a, vandokkum05a, lotz06a,
depropris07a}.  In each case, the estimate of the merger rate proceeds
by estimating the abundance of the class, some time interval over
which they remain identifiably part of that class.  The measurements
of the LRG accretion rate presented here also follow this methodology.
However, our measurements are more reliable than most previous
measurements for a number of reasons.  The first is that the LRGs form
a very uniform, very massive population, as described above, and
therefore dynamical times and dynamical-friction times relevant to
close pairs are straightforward to estimate.  The second is that we
make maximal assumptions so as to put a strict \emph{upper limit} on
the accretion rate.  This is interesting, because the upper limit we
determine is on the low side of existing predictions and measurements.
The third is that we use a technique for measuring the mean number of
close pairs in real space, with no contamination by projected pairs,
so our pre-merger candidate list is clean of such interlopers (in a
statistical sense).

Our technique for measuring the close pairs builds on previous work
\citep{masjedi06a} in which we showed that we could measure projected
correlation functions on extremely small scales ($\kpc$ to $\Mpc$
scales) without the need for complete spectroscopic samples.  We used
this method to overcome the fiber-collision incompleteness of the LRG
sample in the SDSS.  In addition, we showed that the clustering signal
so measured can be deprojected and integrated to deduce close pair
statistics and therefore a rate of merger events among the galaxies in
the sample. We found that LRG--LRG mergers are extremely rare events
and do not play a significant role in the growth of these galaxies, at
least at low redshifts.

In this paper we expand this technique to measure not only the
auto-correlation function of a set of galaxies, but the
cross-correlation function of two different galaxy sets, only one of
which requires spectroscopic information.  We choose LRGs as our
primary spectroscopic sample and we cross-correlate them with distinct
subsamples of companion galaxies over a range of luminosities and
colors.  We convert these results into an accretion rate of
luminosity---within each subsample---into LRGs.  We can estimate a
total accretion rate by this method with unprecedented precision.

Throughout this paper, all distances are comoving, calculated for a
cosmological world model with
$(\Omega_\mathrm{m},\Omega_\Lambda)=(0.3,0.7)$ and Hubble constant
parameterized by $H_0\equiv 100\,h\,\km\,\s^{-1}\,\Mpc^{-1}$.  All
magnitudes are AB.

\section{Data}

The SDSS \citep{stoughton02a,abazajian03a,abazajian04a} has performed
an imaging and spectroscopic survey of $\sim 10^4$ square degrees
\citep{fukugita96a, gunn98a, gunn05a}. Automated, real-time monitoring
\citep{hogg01a}, image processing \citep{lupton01a, stoughton02a,
pier03a}, photometric calibration \citep{smith02a, ivezic04a,
tucker06a, padmanabhan07a}, galaxy target selection for spectroscopy
\citep{strauss02a, eisenstein01a}, design of spectroscopic plates
\citep{blanton03a}, and spectroscopic reductions have produced
enormous, very uniform samples. Of the various SDSS subsamples, the
one that uniformly maps the largest volume is the LRG sample
\citep{eisenstein01a}.

\subsection{Spectroscopic subsample}

The spectroscopic LRG sample is constructed from color-magnitude cuts
in $g$, $r$, and $i$ bands to select galaxies that are likely to be
luminous early-type galaxies at redshifts between 0.15 and 0.5. The
selection is highly efficient and the redshift success rate is
excellent. The sample is constructed to be close to volume-limited up
to $z=0.36$, with a dropoff in density toward $z=0.5$.  Because the
LRG sample is so uniform, and because it occupies such a large volume,
we have used it to demonstrate the homogeneity of the Universe
\citep{hogg05a}, to locate the baryon acoustic feature at low redshift
\citep{eisenstein05b}, and to measure clustering at intermediate and
small scales \citep{zehavi05a, masjedi06a}.

This study uses a spectroscopic sample drawn from NYU LSS {\tt
sample14} \citep{blanton05a}.  This covers 3,836 square degrees and
contains 55,000 LRGs with redshifts $0.16<z<0.47$. The subsample of
LRGs used in this paper has luminosity and redshift ranges of
$-23.2<\MMg <-21.2\,\mag$ and $0.16<z<0.30$.  We 
restrict our sample to $z<0.30$ to allow measurement of
cross-correlations between LRGs and much less luminous companions.
The LRG absolute magnitudes include Galactic extinction corrections
\citep{schlegel98a}, $K$ corrections \citep{blanton06b} and passive
evolution corrections (the latter were applied only for the purposes
of selecting a sample that does not substantially change with
redshift).  These cuts left $23043$ LRGs in our spectroscopic
subsample.

In the SDSS, spectroscopic targets were assigned to spectroscopic
fiber plug plates with a tiling algorithm that ensures nearly complete
samples \citep{blanton03a}. The angular completeness is characterized
for each unique region of overlapping spectroscopic plates
(``sector'') on the sky. An operational constraint of SDSS
spectrographs, however, is that the physical size of the fiber
coupling forces the angular separation of targets on any individual
spectroscopic plate to be larger than $55\,\arcsec$. This ``fiber
collision'' constraint is partly reduced by having roughly
$40\,\percent$ of the sky covered by overlapping plates, but it still
results in $\sim 7\,\percent$ of targeted galaxies not having measured
redshifts.  Because this project involves cross-correlating
spetroscopic and imaging objects, this fiber collision limit only
comes into our analysis in our weighting scheme to account for
incompleteness; it does not affect our pair counts directly.

For each galaxy $j$ in the spectroscopic subsample we compute a weight
$p_j$ that accounts statistically for the spectroscopic incompleteness
coming from fiber collisions.  We calculate this weight by running a
two-dimensional friends-of-friends grouping algorithm on the SDSS
target parent sample in {\tt sample14}, with a $55\,\arcsec$ linking
length.  This procedure emulates the SDSS tiling algorithm
\citep{blanton03a}.  Within each ``collision group'' made by the
friends-of-friends algorithm, we find the number of objects with
measured spectroscopic redshifts and divide by the total number. The
inverse of this ratio is a weight $p_j$ assigned to each spectroscopic
LRG to account for survey incompleteness.

We have created large catalogs of randomly distributed points based on
the SDSS subsample angular and radial (redshift distribution)
models. These catalogs match the redshift distribution of the LRGs and
are isotropic within the survey region.  These catalogs allow us to
check the survey completeness of any given volume and provide a
homogeneous baseline (expected numbers) for the tests that follow.

For each random point $j$ we compute a weight $f_j$ that accounts for
the incompleteness of the spectroscopic survey in that point's region
of the sky \emph{not} due to fiber collision but due to all the other
selection effects in the survey. The {\tt sample14} package provides
the angular geometry of the spectroscopic survey expressed in terms of
spherical polygons. The geometry is complicated: the spectroscopic
plates are circular and overlap, while the imaging is in long strips
on the sky, and there are overlap regions for some plates that have
not yet been observed. The resulting spherical polygons track all
these effects and characterize the geometry in terms of ``sectors'',
each being a unique region of overlapping spectroscopic plates. In
each sector, we count the number of possible targets (LRG, Main, and
Quasar), excluding those missed because of fiber collisions, and the
number of these whose redshifts were determined. We weight the points
in the random catalog matched to the spectroscopic LRGs by the inverse
of the ratio of these numbers ($f_j$). In truth, the priority of all
targets are not equal, such that LRGs always ``lose'' to quasar
candidates, but the LRG priority is equal to that of the dominant MAIN
targets. Only about $12$ percent of the fibers are assigned to
quasars, hence quasar-LRG collisions are rare and this priority bias
is small.

We are required to treat the fiber collision incompleteness $p_j$ and
overall incompleteness $f_j$ factors separately, because the former is
strongly correlated with LRG environment, and there can be physical
differences between LRGs in high and low density environments.

\subsection{Imaging subsamples}

For our imaging data we use the full imaging sample of the SDSS
imaging catalog in the {\tt DR4plus} footprint, which completely
covers {\tt sample14} and is equivalent to the SDSS Data Release
5. After correcting for Galactic extinction, we applied an $i$-band
apparent magnitude cut of $m_i<21\,\mag$. This cut guarantees
completeness in the redshift range $z<0.30$, $K$-corrected absolute
magnitude range $\MMi <-18\,\mag$.  The $\ggz$, $\rrz$, and $\iiz$
bandpasses are the SDSS $g$, $r$, and $i$ bandpasses shifted blueward
by a factor of 1.3 so that $K$ corrections for galaxies at $z=0.3$
become trivial \citep[\eg,][]{hogg02c, blanton03d}.  We applied an
additional surface-brightness cut of $\mu<28\,\mag$ in
$1\,\arcsec^{2}$ in $g$, $r$ and $i$. The surface-brightness cut is
far below the SDSS detection limit; it cleans the data of the obvious
mis-measurements of the Petrosian aperture and other extended-source
data artifacts.

We cannot $K$-correct individual galaxies in the imaging subsample once
and for all, because we do not have spectroscopic redshifts for them,
but each time we consider a \emph{pair} of galaxies, one from the
spectroscopic subsample and one from the imaging subsample, we
fictitiously assign the spectroscopic redshift to the imaging
galaxy. This allows us to calculate for each imaging galaxy in each
spectroscopic--imaging pair a ``temporary'' K-corrected $\iiz$-band
absolute magnitude and $[\ggz-\rrz]$ color for the purposes of
that pair.  We discard these values and compute new ones when the
imaging galaxy is used in another pair with another spectroscopic
galaxy.

We calculated the $K$-corrections using the code {\tt kcorrect}
\citep{blanton06b}. This code is accurate but too slow to calculate the
$K$-corrections individually for the number of pairs ($\sim 10^9$)
found in the cross-correlations. To save time, we computed the
K-correction on a grid of colors in advance. We took galaxies from the
SDSS Main Sample as representative of all galaxy types. We computed
their K-corrections on a grid of redshifts between 0.16 and 0.30 (the
redshift limits of our spectroscopic subsample). We saved the mean
$K$-correction in a grid of observed $[g-r]$ color, $[r-i]$ color, and
redshift. Thereafter we interpolated this cube when calculating the
$K$-correction for an galaxy in any imaging subsample. This speeds up the
$K$-correction procedure immensely and only introduces percent-level
errors in the results.

We have created large catalogs of randomly distributed points, with
the angular distribution of the imaging data subsamples.

\section{Method and results}

As we describe in this section, we cross-correlate galaxies in a
spectroscopic subsample $s$ (of LRGs in this case) with galaxies in an
imaging subsample $i$ to obtain the real-space, projected
cross-correlation function $\wwsi(\rrp)$ as a function of tangential
projected separation $\rrp$.  We de-project this projected
cross-correlation function to obtain the true, three-dimensional,
real-space cross-correlation function $\xisi(r)$ as a function of
real-space separation $r$.  We use this three-dimensional
cross-correlation function and dynamical arguments to place limits on
the accretion rate of objects from subsample $i$ into objects from
subsample $s$, and therefore the mass growth rate of LRGs.

\subsection{Projected cross-correlation function}

In analogy to the definition of auto-correlation function, the
three-dimensional real-space cross-correlation function $\xisi(r)$ of
two subsamples of galaxies $s$ and $i$, is defined as the excess
probability of finding a galaxy from subsample $s$ at a distance $r$
from a galaxy from the subsample $i$, relative to the ``null'' Poisson
prediction.  If we take two small comoving volumes $\di V_s$ and $\di
V_i$, in which we look for galaxies from subsamples $s$ and $i$
respectively, separated by a distance $r$, the expected number of
pairs $\di N_{si}$ with one galaxy coming from subsample $s$ and the
other from subsample $i$ is:
\begin{equation}\label{eq:xis1s2}
 \di N_{si}=\nns\,\nni\,\left[1+\xisi(r)\right]\,\di V_s\,\di V_i \quad ,
\end{equation}
where $\nns$ and $\nni$ are the three-dimensional comoving number
densities of subsamples $s$ and $i$ respectively.

The projected two-dimensional cross-correlation function $\wwsi(\rrp)$
is related to the three-dimensional real-space correlation function
$\xisi(r)$ by a projection over the component $\pi$ of the separation
along the line of sight
\begin{equation}\label{eq:wp}
 \wwsi(\rrp)=\int\di\pi\,\xisi\left(\sqrt{\rrp^2+\pi^2}\right) \quad .
\end{equation}
The two-dimensional function $\wwsi(\rrp)$ has dimensions of length.
Because in practice the correlation function $\xisi(r)$ is very large
at small scales, the integral is dominated by scales $\pi<\rrp$.
Observationally, $\wwsi(\rrp)$ is much more accessible than
$\xisi(r)$, because the radial component of the separation is never
well measured (and not measured at all in the work presented here).

Following the approach we have used previously \citep{masjedi06a}, we
measure $\wwsi(\rrp)$ schematically as a difference of two ratios:
\begin{equation}
\nni\,\wwsi(\rrp)=\frac{\DDs\DDi}{\DDs\RRi}-\frac{\RRs\DDi}{\RRs\RRi} \quad ,
\label{eq:estimator}
\end{equation}
where $\nni$ is the average comoving three-dimensional volume density
of the imaging subsample, the symbols $\DDs$ and $\DDi$ represent the
spectroscopic and imaging data subsamples, and $\RRs$ and $\RRi$
represent the random catalogs matched to the spectroscopic and imaging
subsamples respectively.  The product of a volume density and a
length, $\nni\,\wwsi(\rrp)$ has dimensions of inverse (comoving) area.
In a rough sense, the first term on the right-hand side of equation
(\ref{eq:estimator}) measures the abundance of pairs, and the second
term subtracts the mean background level.  The procedure described
here has been tested with simulations and shown to deliver an unbiased
measure of the correlation function \citep{masjedi06a}.

The main difference between equation~(\ref{eq:estimator}) and our
previous work \citep{masjedi06a} is that the number density $\nni$
that enters on the left-hand side is the number density of the imaging
subsample, which we cannot determine explicitly within this data set
since, by construction, the imaging subsample has no (or few)
spectroscopic redshifts.  We can only measure robustly the product
$\nni\,\wwsi(\rrp)$.  Fortunately, for the purposes of estimating the
merger rate, we need only this product, and not either quantity
separately.

The right-hand side of Equation (\ref{eq:estimator}) includes a
spectroscopic--imaging pair-count factor $\DDs\DDi$:
\begin{equation}
\DDs\DDi=\frac{\displaystyle\sum_{j \in \setDDs\setDDi}p_j}%
              {\displaystyle\sum_{k \in \setDDs}p_k} \quad ,
\label{eq:dsdi}
\end{equation}
where the top sum is over pairs $j$ with one member taken from
the spectroscopic subsample and one from the imaging subsample in some
bin of transverse radii $\rrp$, the bottom sum is over galaxies $k$
from the spectroscopic subsample, and $p_j$ is the weight given to the
spectroscopic galaxy in pair $j$ that accounts for fiber-collision
incompleteness as described above.  Being a sum of dimensionless
weights, this factor $\DDs\DDi$ is dimensionless.

There is a spectroscopic--random pair-count factor $\DDs\RRi$:
\begin{equation}\label{eq:secterm}
\DDs\RRi=\frac{\displaystyle\sum_{j \in \setDDs\setRRi}p_j}%
              {\displaystyle\sum_{k \in \setDDs}p_k
\,\left[\frac{\di \Omega}{\di A}\right]_k\,\frac{\di N}{\di \Omega}} \quad ,
\end{equation}
where the top sum is over pairs $j$ with one member taken from the
spectroscopic subsample and one taken from the random catalog matched
to the imaging subsample, the bottom sum is over galaxies $k$ in the
spectroscopic subsample, $\left[\frac{\di \Omega}{\di A}\right]_k$ is
the inverse square of the comoving distance to the spectroscopic
galaxy $k$, and $\frac{\di N}{\di \Omega}$ is the number density of
the random imaging catalog per solid angle. The product of these two
derivatives gives the average number of random imaging objects per
unit comoving area around each spectroscopic galaxy, so this factor
$\DDs\RRi$ has dimensions of comoving area.

There is a random--imaging pair-count factor $\RRs\DDi$:
\begin{equation}
\RRs\DDi=\frac{\displaystyle\sum_{j \in \setRRs\setDDi}f_j}%
              {\displaystyle\sum_{k \in \setRRs}f_k} \quad ,
\end{equation}
where the top sum is over pairs $j$ with one member taken from the
random catalog matched to the spectroscopic subsample and one taken
from the imaging subsample, the bottom sum is over points in the
random catalog matched to the spectroscopic subsample, and $f_j$ is
the weight given to the random point in pair $j$ that accounts for the
incompleteness of the spectroscopic survey in that point's region of
the sky \emph{not} due to fiber collisions, as described above.  This
factor $\RRs\DDi$ is dimensionless.

There is a random--random pair-count term $\RRs\RRi$:
\begin{equation}
\RRs\RRi=\frac{\displaystyle\sum_{j \in \setRRs\setRRi}f_j}%
              {\displaystyle\sum_{k \in \setRRs}f_k
\,\left[\frac{\di \Omega}{\di A}\right]_k \frac{\di N}{\di \Omega}} \quad ,
\label{eq:rsri}
\end{equation}
similar to the above but for pairs $j$ with one member taken from the
random catalog matched to the spectroscopic subsample and one taken
from the random catalog matched to the imaging subsample.  This factor
$\RRs\RRi$ has dimensions of comoving area.

In measuring the four factors on the right-hand side of
Equation~(\ref{eq:estimator}), we have used $2$ imaging-galaxy color
bins separating the blue and red imaging galaxies in $[\ggz-\rrz]$
color, $20$ imaging-galaxy luminosity bins, which we choose to cover
the range $-24<\MMi <-18\,\mag$ but have roughly the same number of
imaging galaxies in each, and 15 bins in projected transverse
separation $\rrp$ between the LRG and the accompanying galaxy,
covering the range of $0.01<\rrp<8\,\hMpc$ with logarithmic spacing.

In practice, to compute the factors, we bin all the
specroscopic--imaging pairs according to the imaging galaxy color, the
imaging galaxy luminosity, and the comoving projected separation
$\rrp$ of the pair.  As described above, the $K$ corrections and
separations are computed for each pair using the redshift of the
spectroscopic galaxy.  We perform the sums given in
Equations~(\ref{eq:dsdi}) through (\ref{eq:rsri}) in each bin
separately and thereby construct the $\wwsi$ estimator given in
Equation~(\ref{eq:estimator}).

Figures~\ref{fig:wpred} and \ref{fig:wpblue} show the measurements of
$\nni\,\wwsi(\rrp)$ for red and blue companion galaxies respectively.
We have combined the 20 luminosity bins into 5 to simplify the
figures.  The error bars are estimated using jackknife resampling
covariance matrix with 100 subsamples made contiguous and compact on
the sky (based on SDSS ``targetting chunks'') to be as conservative as
possible with regards to correlated calibration and selection
errors. Note that the error bars for each subsample are smallest on
$\kpc$ scales and become larger for both the smaller and larger
scales. On smaller scales this is due to shot noise; the smaller the
separations the fewer the pair counts. On scales larger than a few
$\hkpc$, the errors grow both due to cosmic variance and the fact that
our method becomes more and more vulnerable to interlopers on larger
scales where the clustering power is weaker and background subtraction
is more noisy. These effects generate high correlations among the
errors of different bins, and explains the smoothness of the curves in
Figures~\ref{fig:wpred} and \ref{fig:wpblue} despite the large
uncertainties in each bin.

\subsection{Three-dimensional statistics}

Under the assumption of spherical symmetry, the two-dimensional,
projected cross-correlation function $\wwsi(\rrp)$ can be
``deprojected'' into the three-dimensional, real-space correlation
function $\xisi(r)$:
\begin{equation}
\nni\,\xisi(r) = -\frac{1}{\pi}
\,\int^{\infty}_r\frac{\di\rrp}{\sqrt{\rrp^2-r^2}}
\,\frac{\di\left[\nni\,\wwsi(\rrp)\right]}{\di\rrp}
\quad ,
\end{equation}
where we have kept this deprojection in terms of the measureable
product $\nni\,\wwsi(\rrp)$.

The correlation function can be converted to pair counts; similarly
the cross-correlation function can be converted into the mean number
$\NNi$ of galaxies from a specific imaging subsample $i$ within a given
small three-dimensional separation $\rrf$ of a member of the
spectroscopic subsample $s$:
\begin{equation}
\NNi =  4\,\pi\,\nni\,\int_0^{\rrf}r^2\,\di r\,\left[1+\xisi(r)\right]
\approx 4\,\pi\,\int_0^{\rrf}r^2\,\di r\,\left[\nni\,\xisi(r)\right] \quad ,
\end{equation}
where we have used the fact that at small scales $\xisi\gg 1$ and the
term in brackets in the approximate expression is the quantity we can
de-project from the two-dimensional projected cross-correlation
function $\nni\,\wwsi(\rrp)$. So we can measure the close pair
fraction for every subsample for which we can measure the
cross-correlation function.

\subsection{Merger rate}

Conversion of a pair fraction measurement into a merger rate requires
a time-scale $\ttmerge$ over which the mean galaxy from imaging
subsample $i$ within separation $\rrf$ will merge with the mean LRG
from subsample $s$.  The merger rate estimate $\Gamma_i$ of galaxies
from sample $i$ into galaxies from sample $s$ per galaxy (from $s$)
per unit time is
\begin{equation}
\Gamma_i = \frac{\NNi}{\ttmerge} \quad ,
\end{equation}
and the mean fractional rate of growth of luminosity of a galaxies
from subsample $s$ from accretion of galaxies from subsample $i$ is
\begin{equation}
\frac{1}{\mean{L_s}}\,\left[\frac{\di L_s}{\di t}\right]_i
= \frac{\NNi\,\mean{L_i}}{\ttmerge\,\mean{L_s}}
\quad ,
\end{equation}
where $\mean{L_s}$ is the mean luminosity of galaxies from subsample
$s$ and $\mean{L_i}$ is the mean luminosity of galaxies from subsample
$i$.

The shortest conceivable merger time $\ttmerge$ estimate (which
produces the largest conceivable estimate of the merger rate) is the
orbital time $\ttorbit$.  A more realistic estimate is a time $\ttdyn$
based on dynamical friction.  But in principal all of these times can
be underestimates (and hence any merger rate based on close pairs can
be an overestimate) because there is undoubtedly a large number of
close pairs that will not merge on any short timescale.  In what
follows, we present the orbital time $\ttorbit$ and dynamical friction
time $\ttdyn$ as two options, but then interpret our merger rate
estimates as upper limits.

All of these merger rate estimates depend, in principle, on the radius
$\rrf$ inside of which we have counted close companions.  However,
over the range of interest in Figures~\ref{fig:wpred} and
\ref{fig:wpblue}, $\wwsi(\rrp)$ scales (something) like $\rrp$,
$\xisi(r)$ scales (something) like $r^2$ and $\NNi$ scales (something)
like $\rrf$.  Similarly, both time-scales (orbital and
dynamical-friction) scale like $\rrf$.  For this reason, the inferred
merger and accretion rates (which are based on ratios of $\NNi$ with
the timescales) do \emph{not} depend strongly on the choice of $\rrf$.

The average orbital velocity for a companion around a more massive
galaxy with velocity dispersion $\sigma_v$ is roughly $1.5$ times the
velocity dispersion, so
\begin{equation}
\ttorbit \approx \frac{2\,\pi\,\rrf}{1.5\,\sigma_v} \quad .
\end{equation}
This is the shortest conceivable mean merger time (we have included
the factor of 1.5 to be conservative).  The fractional luminosity
accretion rate estimate for this assumed time-scale is shown with
dashed lines in Figure~\ref{fig:masspec} as a function of the
luminosity of imaging subsample $i$ for red and blue imaging galaxies.
The per-subsample merger rates have been divided by the
absolute-magnitude bin width so that the total fractional accretion
rate is the area under (integral of) the curves.

The Chandrasekhar approximation to dynamical friction is longer than
the dynamical time by a factor roughly equal to the ratio of the mass
of the heavier galaxy to the lighter one. This approximation may
actually be an underestimate of the total merger time found in
explicit $N$-body simulations \citep{boylankolchin07a}, which
serves to strengthen the interpretation of our merger rate estimate as
an upper limit.  For our case, the approximation becomes
\begin{equation}
\ttdyn= \ttorbit\,\frac{\mean{m_s}}{\mean{m_i}} \quad ,
\end{equation}
where $\mean{m_s}$ and $\mean{m_i}$ are the averages of the masses of
the spectroscopic and imaging subsamples respectively, and we have
assumed $\mean{m_s}>\mean{m_i}$.  Keeping things observational, we do
not try to measure masses for galaxies in this work. Instead, we make
the naive assumption that a galaxy's mass is directly proportional to
its $\iiz$-band luminosity and therefore we use the ratio of the
luminosities instead of the masses.  The $\iiz$-band luminosity is
very close to the rest frame $r$-band luminosity. This assumption
works fairly well for the mass ratios of red imaging galaxies to the
spectroscopic galaxies (which are LRGs) but tends to over estimate the
masses of the blue companions. This bias leads to an underestimation
of merger time-scales and hence an overestimation of the fractional
accretion rate for blue companions.

The solid lines in Figure~\ref{fig:masspec} show the calculated
fractional luminosity growth of the LRGs assuming the dynamical
friction time-scale $\ttdyn$ for the mergers.  The per-subsample
merger rates have been divided by the absolute-magnitude bin width so
that the total fractional accretion rate is the area under (integral
of) the curves.  If the results in Figure~\ref{fig:masspec} are
naively interpreted as fractional \emph{mass} accretion rates (they
are fractional luminosity rates), the blue galaxies are doubly
overestimated, because both the merger rate (inverse timescale) and
the delivered mass have been over-estimated.

Under the orbital time-scale assumption, the growth curve in
Figure~\ref{fig:masspec} peaks near the magnitude of $\Lstar$
galaxies. This represents fact that most of the light in the Universe,
even near LRGs, is in $\Lstar$ galaxies. Under the dynamical
friction assumption, the curves shift to more luminous, more massive,
galaxies; lighter galaxies linger around the LRG for a longer time.

The maximal fractional luminosity accretion rate (the sum of the
integrals under the dashed curves in Figure~\ref{fig:masspec}) is
$[5.6\pm 0.2]\,\percent\,\Gyr^{-1}$, but this rate is unrealistically
high; certainly pairs with large mass differences do not merge in an
orbital time!  The dynamical-friction rate is $[1.7\pm
0.1]\,\percent\,\Gyr^{-1}$, and is also probably an overestimate
because at least some physical pairs are not on the path to rapid
merging.

Table~\ref{tab:data1} gives the derived fractional luminosity growth
from every imaging subsample for both the orbital time-scale
assumption and the dynamical friction time-scale assumption as a
function of the luminosity and color of the subsamples.

\section{Discussion}

We have combined Sloan Digital Sky Survey (SDSS) spectroscopic data on
$23043$ luminous red galaxies (LRGs) with SDSS imaging data on
enormous subsamples of fainter galaxies to measure cross-correlations.
We have measured the projected two-dimensional cross-correlation
functions $\wwsi(\rrp)$ on very small scales ($0.01<\rrp<8\,\hMpc$)
between spectroscopic LRGs (``$s$'') with luminosities $\MMi\approx
-22.75\,\mag$ and many subsamples of imaging galaxies (``$i$'') with
luminosities $\MMi <-18\,\mag$. The imaging limit
is $50$ times or $4.25\,\mag$ fainter than the mean LRG; the samples
of companion galaxies cover a broad range in color and magnitude. In
addition, the large volume of the SDSS LRG sample allows us to cut the
companion galaxies into many distinct subsamples with different
luminosities and colors but nonetheless measure the clustering as a
function of these properties with high signal-to-noise. The principal
limitation arises from the lack of spectroscopic information on the
companion galaxies; this makes it impossible to precisely measure the
real-space number densities for the companion subsamples.  We cannot
disentangle the clustering power from the number density; we only
measure the product $\nni\,\wwsi(\rrp)$ but not either $\nni$ or
$\,\wwsi(\rrp)$ separately.

\begin{figure}
\begin{center}
\includegraphics[width=1\textwidth]{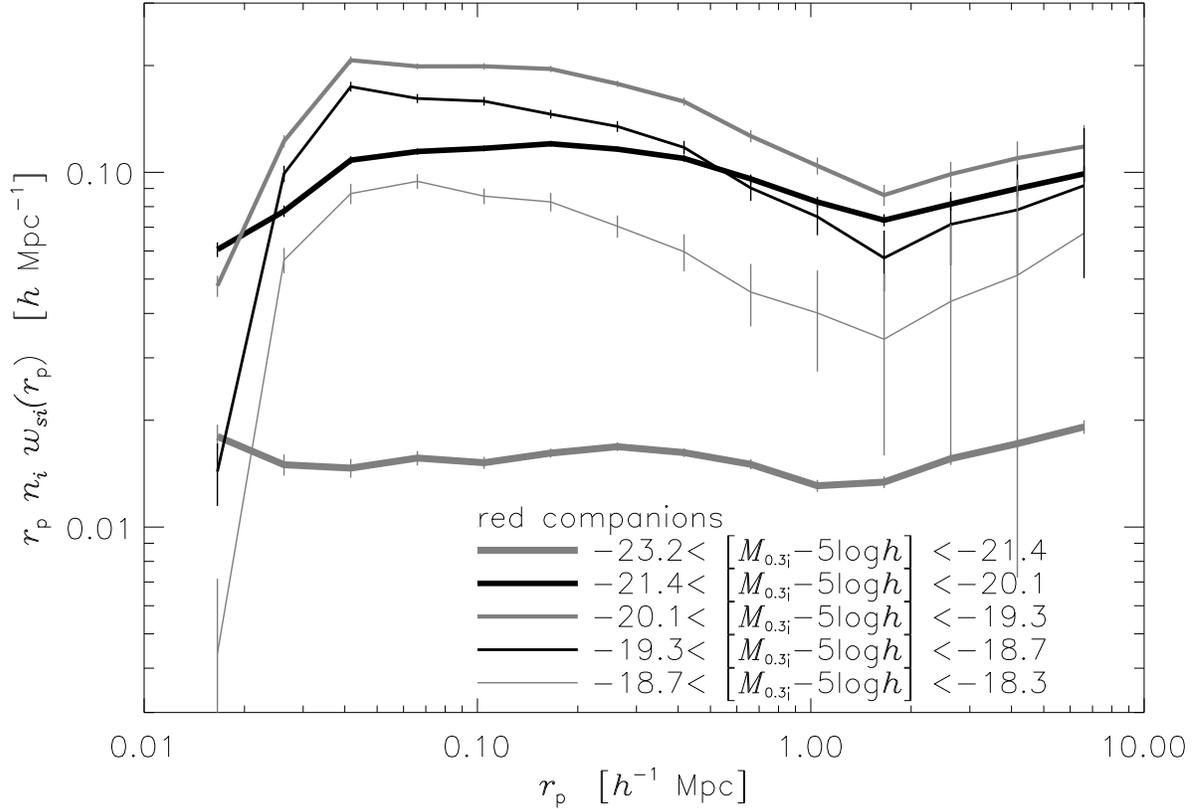}
\caption{Projected two-dimensional cross-correlation functions
 $\nni\,\wwsi(\rrp)$ of spectroscopic LRGs $s$ with \emph{red}
 companion imaging galaxy subsamples $i$ with different luminosity
 ranges, weighted by the number density of each imaging subsample,
 $\nni$, and scaled by $\rrp$ for better illustration.  The
 spectroscopic LRG subsample $s$ has absolute magnitudes $-23.2<\MMg
 <-21.2\,\mag$ and has been trimmed to redshifts $0.16<z<0.30$. The
 uncertainties are estimated by jackknife, with jackknife trials
 dropping contiguous sky regions (see text).}
\label{fig:wpred}
\end{center}
\end{figure}

\begin{figure}
\begin{center}
\includegraphics[width=1\textwidth]{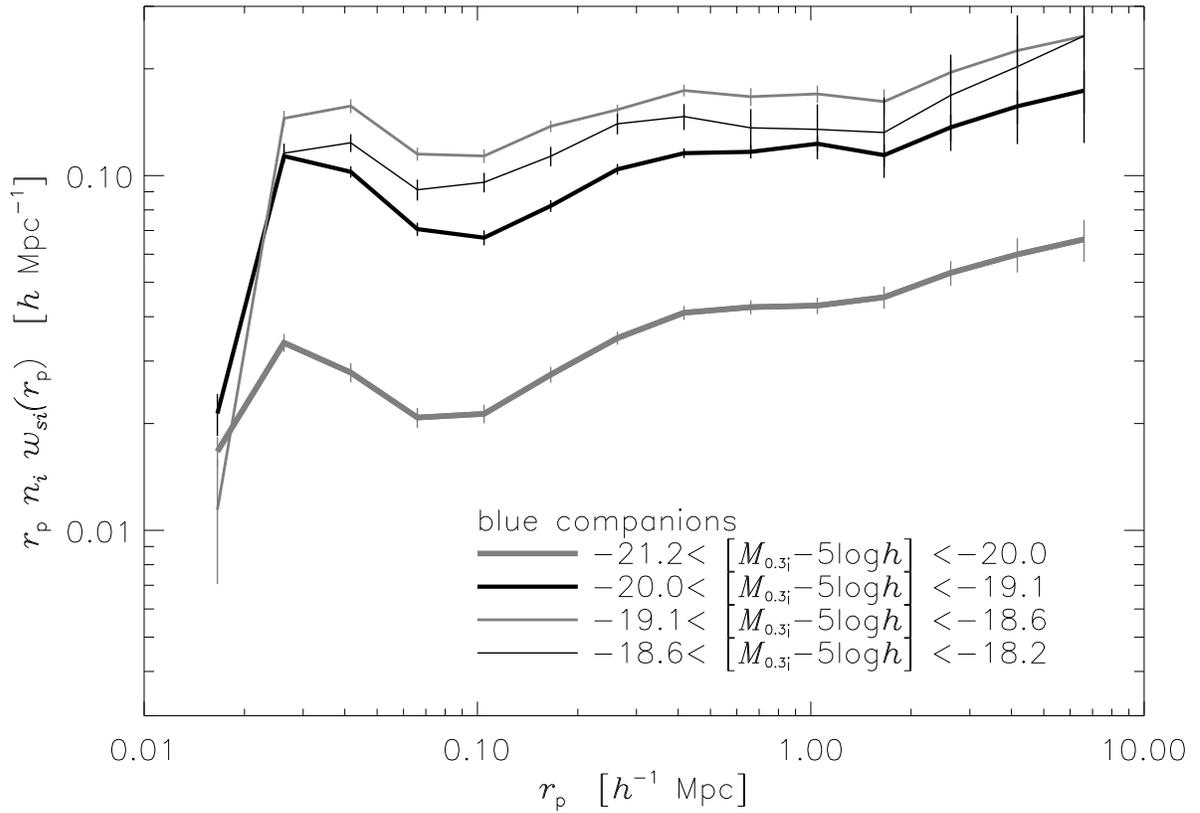}
\caption{Same as Figure \ref{fig:wpred} but for \emph{blue} companion
imaging galaxy subsaples $i$.}
\label{fig:wpblue}
\end{center}
\end{figure}

Figures~\ref{fig:wpred} and \ref{fig:wpblue} show the results of these
measurements of $\nni\,\wwsi(\rrp)$ for red and blue galaxies
respectively. In both figures several characteristic transition scales
are visible.  The sharp break at at $\rrp\approx 2\,\hMpc$ and the
less-sharp transition at $\rrp\approx 0.3\,\hMpc$ in the curves can be
explained in the context of the ``halo occupation'' picture of galaxy
clustering \citep{peacock00a, scoccimarro01a, berlind02a}.  If
galaxies are residing within dark matter halos then the clustering of
the galaxies on scales larger than halos is determined by the
clustering of the dark matter halos that host them, plus statistics of
the occupation of halos by galaxies.  In this picture, the first
transition at $\rrp\approx 2\,\hMpc$ locates the size of the largest
halos that host LRGs---the largest halos in the Universe.  At larger
separations, at $\rrp> 2\,\hMpc$, this is the regime in which all
LRG--galaxy pairs come from two separate halos (the ``two-halo''
regime).  Inside this scale, at $0.3<\rrp<2\,\hMpc$, the galaxy--LRG
pairs are a mix of pairs, in some of which the companion galaxy
belongs to the same halo as the LRG and in some of which the galaxy
belongs to a separate halo.  This ``mixed'' regime comes from the fact
that LRGs reside in a range of halo sizes.  The inner limit of this
regime is at $\rrp\approx 0.3\,\hMpc$, depending on the luminosity of
the imaging galaxy subsample $i$ in question. This inner scale is
close to the virial size of the smallest halo that can host an LRG
(plus the virial size of the smallest halo that can host a companion
galaxy from the imaging subsample $i$). At smaller scales, at $\rrp<
0.3\,\hMpc$, all the galaxies belong to the same halo as the LRG halo
(the ``one-halo'' regime).  Here the clustering represents the mean
radial profile of the halos mixed with details of galaxy evolution.
Figure~\ref{fig:wpred} shows an increase in the clustering of blue
galaxies at $\rrp\approx 50\,\hkpc$ toward the central regions of the
halo.  This could be caused by a boost of star formation in these
galaxies, which makes them more luminous and places them in higher
luminosity bins in our calculation.

Finally, both Figures show a sharp drop in the clustering power on
scales $\rrp<30\,\hkpc$.  This could be due to failure of the object
detection software of the SDSS \citep[\eg,][]{masjedi06a}, or it could
be a real effect from disintegration of galaxies by dynamical friction
or other tidal stripping expected in some galaxy evolution models.

We integrate the de-projected, three-dimensional cross-correlation
functions $\nni\,\xisi(r)$ for each imaging subsample $i$ on very
small scales to calculate the average number of galaxies that are in
dynamical pairs with each LRG. We use two different time-scales for
merger events to calculate merger rates.  The first is the orbital
time $\ttorbit$, equivalent to assuming that all galaxies merge in one
orbit. This is the shortest time imaginable to merge, so it provides a
strict upper limit on the merger rate.  The second time-scale is the
dynamical friction time-scale $\ttdyn$ for which we approximate the
merger time with a linear function of the mass ratio
$\mean{m_s}/\mean{m_i}$ of the mean galaxies from the two samples.
This is equivalent to assuming that equal-mass (LRG--LRG) mergers take
the one-orbit time, but pairs of galaxies with more different masses
take longer times to merge.

\begin{figure}
\begin{center}
\includegraphics[width=1\textwidth]{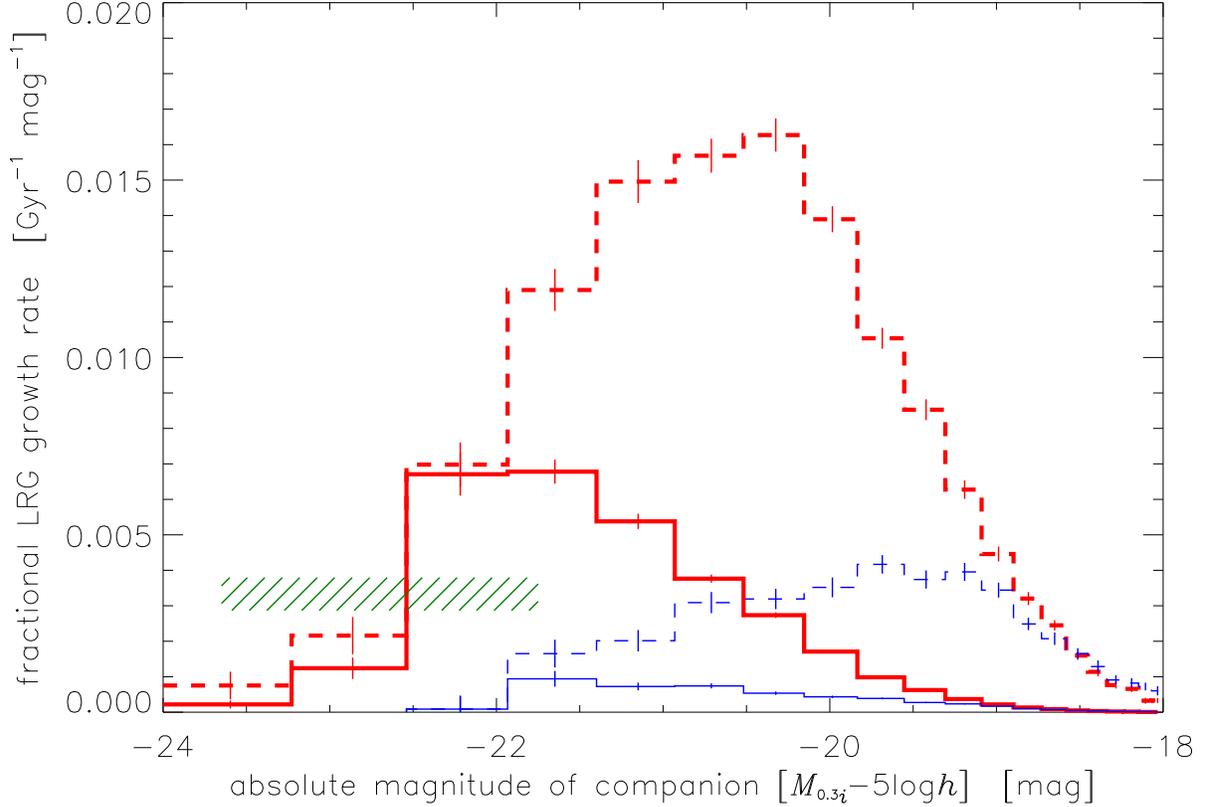}
\caption{The mean fractional luminosity growth of LRGs per $\Gyr$ per
unit absolute magnitude of the companion, derived from the data in
Table~\ref{tab:data1}. The dashed thick red and thin blue lines show
this quantity for the red and blue companions respectively, made under
the maximal assumption that all companion galaxies merge into the LRG
in one orbital time.  The solid thick red and thin blue lines show the
same thing but made under the more realistic assumption that companion
galaxies merge in a time governed by dynamical friction.  The total
fractional growth rate is the integral (area) under the curves.  The
green hatched region shows the result of our previous work
\citep{masjedi06a}.}
\label{fig:masspec}
\end{center}
\end{figure}

We use the two time-scales to calculate both a strict upper limit to
the merger rate and a more realistic rate, although even the dynamical
friction time-scale calculation involves assuming that essentially all
close pairs merge.  For both time-scales we have measured the fraction
of LRG luminosity that is added to the LRG through mergers of galaxies
from each imaging subsample $i$ per $\Gyr$.  The fractional luminosity
growth for LRGs through mergers as a function of the color and
magnitude of the merging companions is shown in
Figure~\ref{fig:masspec}.

Most of the luminosity brought into LRGs by merging is brought by red
companions or ``dry mergers,'' and most of it is brought by galaxies
near (or above) $\Lstar$.  The contribution to growth decreases with
decreasing luminosity at the faint end; the curves essentially to zero
by $\MMi =-18\,\mag$.  Calculation of the total amount of luminosity
brought in by merger activities does not require consideration of
fainter companion galaxies.

Integration of Figure~\ref{fig:masspec} over companion absolute
magnitude yields the total fractional amount of LRG luminosity growth
from mergers.  Under the maximal one-orbit merger time assumption, we
find that the total fractional growth rate is strictly smaller than
$[5.6\pm 0.2]\,\percent\,\Gyr^{-1}$.  Under the more realistic
dynamical friction assumption, we find that the total fractional
growth rate is $[1.7\pm 0.1]\,\percent\,\Gyr^{-1}$ where only about
one-tenth of that is through ``wet mergers'' (blue companions) and the
rest is through dry mergers (red companions).

Our results are \emph{not} consistent, at face value, with most
morphological measurements of the merger rate---measures that involve
identification of merging galaxies by their appearances---most of
which find rates on the order of ten~percent per Gyr or of order unity
over a Hubble time (\eg, \citealt{abraham96, conselice03a,
vandokkum05a, lotz06a}; though see also \citealt{depropris07a}).  Our
method for inferring the merger rate suffers---as all these other
investigators' methods do---from uncertainties in merger timescales.
However, we avoid all issues related to morphological selection of
merging systems or merger remnants, which tend to introduce
subjectivity, and we avoid uncertainties related to line-of-sight
projections because we work with the true three-space correlation
function.  Our method is therefore much more precise than other
methods.  Furthermore, because merging cannot happen on timescales
shorter than a dynamical time, our upper limit is extremely robust
(and not made uncertain by projection effects).

Our merger rate estimate can be reconciled with other estimates if we
assume that either \textsl{(1)}~the merger rate is an extremely strong
function of the primary galaxy mass (since we only investigate the
rate for the most massive galaxies in the Universe), or
\textsl{(2)}~merging produces observable distortions to galaxy
morphologies (\eg, tidal tails) that last for many dynamical times, or
\textsl{(3)}~significant morphological signs of merging can be raised
by very frequent, very minor mergers, which don't contribute much to
the build-up of mass.  Our results are more consistent with measures
of the merger rate based on counts of close pairs \citep{carlberg94a,
patton97a, vandokkum99a, carlberg00a, patton00a, lin04a, masjedi06a,
bell06a}, but even there, our results are on the low side.

Our results are also lower than any accretion or merger predicted in
theories of galaxy formation in a cosmological context
\citep{murali02a, maller06a, conroy07a}, but we caution that no
predictions have been made for exactly what we have observed, and that
galaxy--galaxy merging occurs at length and dynamical scales where
cosmological simulations are not completely reliable.

There are three respects in which the luminosity growth shown in the
solid lines in Figure~\ref{fig:masspec}---the more ``realistic''
estimates---are nonetheless over-estimates or upper limits on the true
fractional mass growth for LRGs.  First, we are assuming that the vast
majority of pairs do merge as quickly as dynamical friction allows.
This is not true for close pairs in high velocity-dispersion
environments.  In addition, even when the pairs are bound the
Chandrasekhar formula may be an overestimate \citep{boylankolchin07a}.
Second, the blue galaxies have both their masses over-estimated and
their dynamical friction times under-estimated with a constant
mass-to-light ratio assumption, so the blue galaxies do not contribute
as much mass as Figure~\ref{fig:masspec} implies.  Third, we are
assuming that all the stars in the companion galaxies will end up in
the central LRGs. Recent work has suggested that this is not the case
and in fact up to $50$ percent of the stars in the companions could be
stripped off the companion before the merger is complete and
contribute to the intra-cluster light instead of the luminosity of the
LRG \citep{lin04a}. This suggests that even our dynamical friction
assumption could still be an upper limit on the growth of the LRGs.

These results are consistent with recent results on the evolution of
the luminosity function of the red galaxies since redshift $z\sim 1$,
which find modest evolution \citep{bell04a, blanton06a, wake06a,
faber07a, brown07a}.  If we take our results at face value and assume
that the growth happens at a non-evolving rate, we expect the LRGs to
grow by about $\approx 10$ percent between redshift $z=1$ and $z=0.1$
(a period of $\approx 6\,\Gyr$).

\acknowledgments It is a pleasure to thank Eric Bell, Andreas Berlind,
Daniel Eisenstein, and Ari Maller for valuable input.  Some of this
research was performed while DWH was generously hosted by Hans-Walter
Rix and the Max-Planck-Institut f\"ur Astronomie.  This research was
partially supported by the National Aeronautics and Space
Administration (NASA; grant NAG5-11669) and the National Science
Foundation (NSF; grant AST-0428465).  This research made use of the
NASA Astrophysics Data System.  It also made use of the ``idlutils''
codebase maintained by David Schlegel, Wayne Landsman, Doug
Finkbeiner, and others.

This research made use of public SDSS data.  Funding for the SDSS and
SDSS-II has been provided by the Alfred P. Sloan Foundation, the
Participating Institutions, the National Science Foundation, the
U.S. Department of Energy, the National Aeronautics and Space
Administration, the Japanese Monbukagakusho, the Max Planck Society,
and the Higher Education Funding Council for England. The SDSS Web
Site is http://www.sdss.org/.

The SDSS is managed by the Astrophysical Research Consortium for the
Participating Institutions. The Participating Institutions are the
American Museum of Natural History, Astrophysical Institute Potsdam,
University of Basel, University of Cambridge, Case Western Reserve
University, University of Chicago, Drexel University, Fermilab, the 
Institute for Advanced Study, the Japan Participation Group, Johns
Hopkins University, the Joint Institute for Nuclear Astrophysics, the
Kavli Institute for Particle Astrophysics and Cosmology, the Korean
Scientist Group, the Chinese Academy of Sciences, Los Alamos National
Laboratory, the Max-Planck-Institute for Astronomy, the
Max-Planck-Institute for Astrophysics, New Mexico State University,
Ohio State University, University of Pittsburgh, University of
Portsmouth, Princeton University, the United States Naval Observatory,
and the University of Washington.

\clearpage 
\begin{table}
\begin{center}
{Fractional luminosity growth of LRGs from mergers}
\begin{tabular}{ccccc}
\hline \hline
$\MMi$ & Red max & Blue max & Red DF & Blue DF \\
{}[$\mag$] & [$10^{-3}\,\Gyr^{-1}$] & [$10^{-3}\,\Gyr^{-1}$]
           & [$10^{-4}\,\Gyr^{-1}$] & [$10^{-4}\,\Gyr^{-1}$] \\
\hline
$-24.00$ to $-23.22 $ & $ 0.59\pm0.31 $ & $ 0.35\pm0.52 $ & $ 1.72\pm0.89 $ & $ 1.01\pm1.52$\\
$-23.22$ to $-22.53 $ & $ 1.49\pm0.36 $ & $ 0.02\pm0.33 $ & $ 8.53\pm2.08 $ & $ 0.12\pm1.88$\\
$-22.53$ to $-21.92 $ & $ 4.22\pm0.38 $ & $ 0.05\pm0.23 $ & $ 40.6\pm3.65 $ & $ 0.53\pm2.20$\\
$-21.92$ to $-21.39 $ & $ 6.34\pm0.32 $ & $ 0.88\pm0.21 $ & $ 36.1\pm1.80 $ & $ 5.01\pm1.19$\\
$-21.39$ to $-20.92 $ & $ 7.01\pm0.28 $ & $ 0.94\pm0.14 $ & $ 25.2\pm1.01 $ & $ 3.39\pm0.51$\\
$-20.92$ to $-20.51 $ & $ 6.46\pm0.20 $ & $ 1.27\pm0.12 $ & $ 15.5\pm0.47 $ & $ 3.05\pm0.30$\\
$-20.51$ to $-20.15 $ & $ 5.90\pm0.17 $ & $ 1.16\pm0.10 $ & $ 9.91\pm0.28 $ & $ 1.94\pm0.17$\\
$-20.15$ to $-19.83 $ & $ 4.43\pm0.12 $ & $ 1.12\pm0.09 $ & $ 5.45\pm0.14 $ & $ 1.38\pm0.11$\\
$-19.83$ to $-19.55 $ & $ 2.96\pm0.08 $ & $ 1.17\pm0.07 $ & $ 2.76\pm0.08 $ & $ 1.09\pm0.07$\\
$-19.55$ to $-19.30 $ & $ 2.10\pm0.07 $ & $ 0.92\pm0.06 $ & $ 1.54\pm0.05 $ & $ 0.68\pm0.05$\\
$-19.30$ to $-19.09 $ & $ 1.36\pm0.06 $ & $ 0.86\pm0.05 $ & $ 0.81\pm0.03 $ & $ 0.51\pm0.03$\\
$-19.09$ to $-18.89 $ & $ 0.85\pm0.04 $ & $ 0.66\pm0.04 $ & $ 0.42\pm0.02 $ & $ 0.32\pm0.02$\\
$-18.89$ to $-18.73 $ & $ 0.54\pm0.03 $ & $ 0.42\pm0.03 $ & $ 0.22\pm0.01 $ & $ 0.17\pm0.01$\\
$-18.73$ to $-18.58 $ & $ 0.36\pm0.02 $ & $ 0.31\pm0.03 $ & $ 0.13\pm0.01 $ & $ 0.11\pm0.01$\\
$-18.58$ to $-18.45 $ & $ 0.21\pm0.02 $ & $ 0.21\pm0.02 $ & $ .066\pm.005 $ & $ .068\pm.006$\\
$-18.45$ to $-18.34 $ & $ 0.13\pm0.01 $ & $ 0.15\pm0.02 $ & $ .037\pm.004 $ & $ .042\pm.006$\\
$-18.34$ to $-18.23 $ & $ 0.08\pm0.01 $ & $ 0.09\pm0.01 $ & $ .020\pm.003 $ & $ .023\pm.003$\\
$-18.23$ to $-18.15 $ & $ .058\pm.008 $ & $ .073\pm.012 $ & $ .014\pm.002 $ & $ .017\pm.003$\\
$-18.15$ to $-18.07 $ & $ .025\pm.006 $ & $ .056\pm.008 $ & $ .005\pm.001 $ & $ .012\pm.002$\\
$-18.07$ to $-18.00 $ & $ .023\pm.005 $ & $ .041\pm.009 $ & $ .005\pm.001 $ & $ .008\pm.002$\\
\hline
\end{tabular}
\caption[Measurements of the fractional growth of the LRGs over a
$\Gyr$ through merger events.]{\label{tab:data1} Measurements of the
fractional growth of spectroscopic LRGs over a $\Gyr$ through merger
events with different imaging subsamples of companion galaxies with
different magnitude ranges and colors.  The second and third (``max'')
columns present the measurements under the maximal assumption that all
pairs merge in an orbital time.  The fourth and fifth (``DF'') columns
present the measurements under the more realistic dynamical-friction
assumption.  Note that the ``max'' and ``DF'' columns are given in
units that differ by a factor of 10.}
\end{center}
\end{table}

\clearpage 

\clearpage 

\clearpage

\end{document}